\documentclass[aps,prc,twocolumn]{revtex4}
\usepackage{graphicx}
\def\be{\begin{equation}}
\def\ee{\end{equation}}
\begin{document}
\title{Moment distributions of clusters and molecules in the 
adiabatic rotor model}
\author{ G. E. Ballentine$^1$, G. F. Bertsch$^2$, N. Onishi$^3$, K. Yabana$^4$
}
\address{$^1$Chemistry Division, Argonne National Laboratory, 9700 S Cass
Ave.,
   Argonne, IL 60439 USA\\
  $^2$Institute for Nuclear Theory, University of Washington, Seattle, WA
   98195 USA\\
   $^3$Tokyo International University, 1-13-1 Matoba-kita, Kawagoe, Saitama 350-1197, Japan\\
   $^4$Center for Computational Sciences and Institute of Physics, University of Tsukuba,
   Tsukuba 305-8571, Japan
   }

\begin{abstract}
  We present a Fortran program to compute the distribution of
dipole moments of free particles for use in analyzing molecular
beams experiments that measure moments by deflection in an
inhomogeneous field.   The theory is the same for magnetic and electric
dipole moments, and is based on a thermal ensemble of classical 
particles that are free to rotate and that have moment vectors
aligned along a principal axis of rotation.  The theory has
two parameters, the ratio of the magnetic (or electric)
dipole energy to the thermal energy, and the ratio of moments
of inertia of the rotor.
\end{abstract}
\maketitle
\section{Introduction}
  It is common to measure the magnetic moment or the electric dipole 
moment of clusters or small molecules by the deflection of 
a molecular beam by an inhomogeneous field[1-5].  These experiments take 
place in the gas phase using Stern-Gerlach
magnets to deflect the beam in the magnetic case[4-5] and using an 
electric field gradient in 
the electric case[1-3]. One needs a theory of the
moment distribution to relate the observed deflections to the
intrinsic moment of the particles.  
There are two well-known limits for the distribution, the quantum
limit of a spin with a fixed total angular momentum, and superparamagnetic
limit, where the moments are thermally distributed.
Neither of
these limits is valid for the typical situation of a nanoparticle,
which may have a moment with fixed orientation in a body-centered
frame, but changing orientation in
the laboratory frame.  If we assume that an external field is
introduced adiabatically, the distribution of moments can be
computed using the adiabatic invariants of the rigid rotor.
This classical theory and a method of solution was given in ref.
\cite{be95}. While conceptually the theory is quite straightforward, the 
computation is not completely trivial.  We first present the equations
that govern the deflections, and then the computational aspects.
\section{Theory}
  We consider a ferromagnetic particle having a 
magnetic moment aligned along
the 3-axis and equal moments of inertia around the 1 and 2 axes in
the body-fixed frame.  
Its Lagrangian is given by
\be
L= {J_1\over 2} \left( \dot\theta^2+ \dot\phi^2 \sin^2 \theta\right) + 
{J_3\over 2} \left(\dot\psi + \dot\phi \cos \theta\right)^2 + \mu_0 B
\cos\theta
\ee
where $J_1, J_2=J_1$ and $J_3$ are the principal moments of inertia and 
$\theta,\phi,\psi$ are the Eulerian angles of the 3-axis with respect
to the magnetic field $B$.  The theory would be the same for a particle
with an intrinsic electric dipole moment $p_0$ in an electric field
${E}$, simply replacing $\mu_0 B$ by $p_0 { E}$ in all 
equations.  There are three constants of motion for the 
Lagrangian  eq. (1).   They are the energy $E$,
\be
E= {J_1\over 2} \left( \dot\theta^2+ \dot\phi^2 \sin^2 \theta\right) + 
{J_3\over 2} \left(\dot\psi + \dot\phi \cos \theta\right)^2 - \mu_0 B
\cos\theta
\ee
the angular momentum about the field direction $m_z$,
\be
m_z \equiv {\partial L \over \partial \dot \phi} =
J_1 \dot\phi \sin^2\theta + J_3\left( \dot\psi + \dot\phi \cos\theta\right)
\cos\theta 
\ee
and the angular momentum about the 3-axis $m_3$,
\be
m_3 \equiv {\partial L \over \partial \dot \psi} =
 J_3\left( \dot\psi + \dot\phi \cos\theta\right)
.
\ee
The last quantity, $m_3$, is only conserved because of the condition
we imposed that $J_2=J_1$.  
Under the equation of
motion, the variable $\theta$ 
has a periodic dependence on time, oscillating between two limits
$\theta_1$ and $\theta_2$.  For convenience below, we replace the 
variable $\theta$ by its cosine, $u = \cos\theta$. 
The quantity of interest for the deflection
measurement is the average moment of the particle $\bar\mu
=\mu_0 \bar u$ where $\bar u$ is the average of $u$ over a cycle.
There is an analytic formula for this quantity in terms of 
the elliptic integrals $K(\nu)$ and $E(\nu)$ which can be
compactly expressed in term of the three zeros $u_0,u_1,u_2$
of the cubic polynomial
\be
f(u)=(2J_1 E -J_1 m_3^2/J_3 +2J_1 \mu_0 B u)(1-u^2)-(m_z-m_3u)^2
\ee
The formula for $\bar u$ is\cite{typo}
\be
\label{baru}
\bar u = { u_0 K(\nu) +(u_2-u_0) E(\nu)\over K(\nu)}
\ee
where $\nu = (u_2-u_1)/(u_2-u_0)$.  

We use eq. (\ref{baru}) to compute $\bar u$ as a function of $m_3,m_z$
and $E$.  However, $E$ changes as the particle enters the field.  
Assuming the field change is adiabatic, the action $J_\theta$ associated
with the variable $\theta$ remains constant and thus can be used to
determine the new value of $E$.
There is no analytic expression for $E(J_\theta)$ or even for the
inverse function $J_\theta(E)$.  In the program we compute the
latter from its definition
\be
J_\theta = 2 \int_{u_1}^{u_2} {\sqrt{f(u)}\over 1-u^2} du.
\ee
In zero external field, the action is simply related to the
total angular momentum $I$,
\be
I = \max( |m_3|,|m_z|) + {J_\theta \over 2\pi}.
\ee  
This relation is useful to make a connection to the quantum mechanical
formulation of the problem as well as to make tests of the program.  

The probability distribution $P(u)$ that we seek to compute can now be
expressed as the three-dimensional integral,
$$
P(u) =
$$ 
\be
\label{prob}
{1\over Z(T)} \int_0^\infty dI \int_{-I}^I d m_z
\int_{-I}^I d m_3 \,\,\delta(u - \bar u(I,m_z,m_3)) e^{-E_0/kT} 
\ee
where the partition function $Z(T)$ is the corresponding integral
without the delta function and $E_0 =  (I^2-m_3^2)/2 J_1 + m_3^2/ 2
J_3$.  There are two
symmetries that can be used to reduce the size of the integration 
region.  Namely, $\bar u(I,m_z,m_3) $ remains the same under the interchange
of $m_z$ and $m_3$ and under the replacement $m_3, m_z \rightarrow
-m_3,-m_z $.   While eq. (\ref{prob})
is expressed in terms of dimensioned physical parameters, in fact
the results only depend on two dimensionless combinations of those
parameters, namely
\be
x= {\mu_0 B \over k T}
\ee   
and $J_1/J_3$  Note that the distribution function is independent
of the overall magnitudes of the moments of inertia.

\section{Numerical}
We evaluate the integral (\ref{prob}) using uniform meshes in
the three integration variables, binning values of $\bar u$ on
the mesh points to construct the probability density.  This requires
a fine integration mesh due to the singularities and discontinuities
in the integrand.  We use a mesh size of $\Delta m / I \approx 0.005$ 
to achieve an accuracy suitable for graphing the distribution $P(u)$.
It also helps to have incommensurate mesh spacings for two $m$ 
integrations.

Another numerical problem is connected with determining $\bar u$ as 
a function of $J_\theta$.  Both quantities are computed directly
in terms of the energy variable $E$, but to find $\bar u$ as a function
of $J_\theta$ requires solving an implicit equation.  In the program
this is carried out by Newton's method; a warning is given if the convergence
is poor.
\section{Tests and program use }
There are two analytic tests that can be made of the program.  The
first is the probability distribution at zero field, which is given
\cite{be94}
by
\be
\label{log}
P(u) = {1\over 2} \log(1/|u|).
\ee
Unfortunately, eq. (6) can not be used at $B=0$ because $u_0$
goes to infinity at that point.  
However, the numerical parameters in the program have been set so
that the distributions are accurate to within a few percent
for values of $x$ greate than 0.01.  Fig. 1 show the comparison of eq.
(\ref{log})  with the computed distribution at $x=0.1$ with the mesh 
as given above.  The small irregularities are the binning effects
associated with the finite mesh size.

\begin{figure}
\includegraphics [width = 9cm]{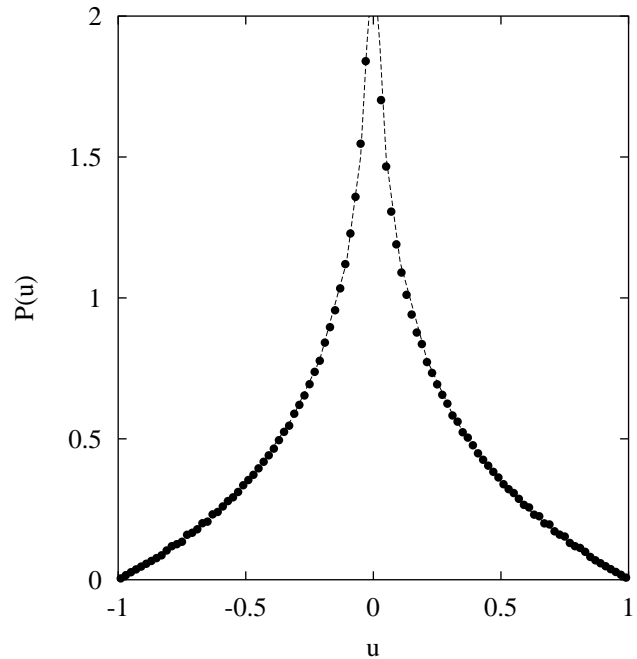}
\caption{Zero field distribution compared with the numerical results
for $x=0.01$ and $J_1/J_3=1$.
}
\label{M_eig}   
%fig 1 see /scratch/refit/plot/eigen2.gnu
\end{figure} 

The second analytic test is the ensemble-average moment 
$\langle \bar u \rangle$
at small fields.  It is given by
\be
\label{2over9}
 \langle \bar u \rangle \approx {2\over 9} x
\ee
The computed ensemble average for $x=0.01$ is $ \langle \bar u \rangle =
0.000222 $, in excellent agreement with eq. (\ref{2over9}).

The program runs without any input file, as all of the parameters
have been set in the Fortran coding.  The important physical parameters
$x$ and $J_1/J_3$ are specified on lines 22 and 25 of the code,
respective.  Running the code with the values given,
{\tt
\begin{verbatim}
      betamu0B=1.0d0
      J1J3=1.d0
\end{verbatim}
}
gives as direct output the values of $x,J_3/J_1$ and $ \langle \bar u
\rangle $, 
\begin{verbatim}
   1.000   1.000   0.19220
\end{verbatim}
The program also writes a data file `udist.dat' that has a table of
values of $u$ and $P(u)$.  Fig. 2 shows a plot of that data.
\begin{figure}
\includegraphics [width = 9cm]{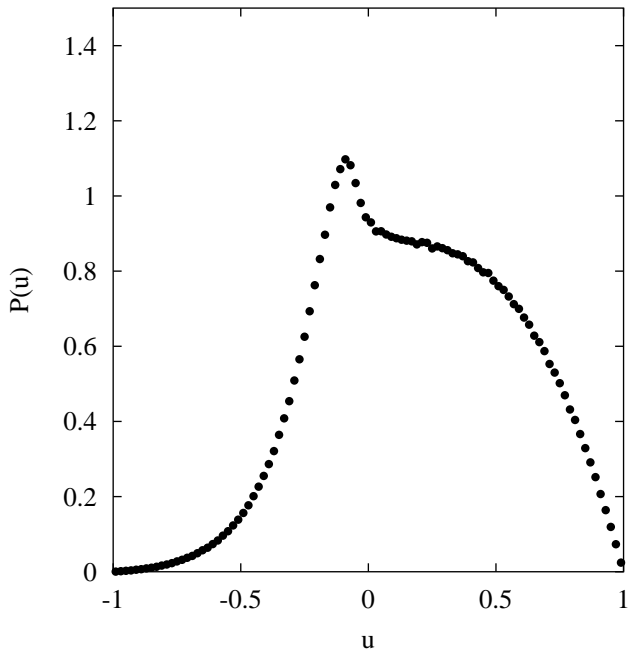}
\caption{Distribution $P(u)$ computed for $x=1$ and $J_1/J_3=1$.
}
\end{figure} 

\section{Acknowledgment}

This work is supported by the US Department of Energy, 
Office of Basic Energy Sciences, Division of Chemical Science, 
under Contract No. W-31-109-ENG-38 and by the Office of Nuclear Physics
under Grant DE-FG02-00ER41132 and by CREST (Core Research for Evolutional Science and Technology) of Japan Science and Technology Agency (JST).


\begin{thebibliography}{00}
\bibitem{kn04}
M.B. Knickelbein, J. Chem. Phys., 120, 10450 (2004)
\bibitem{mo03}
R. Moro, X. Xu, S. Yin, and W. A. de Heer, Science 300, 1265 (2003)
\bibitem{kn01}
M. B. Knickelbein, J. Chem Phys., 115, 5957 (2001)

\bibitem{kn01b}
M. B. Knickelbein, Phys. Rev. Lett., 86, 5255 (2001)

\bibitem{co93}
A. J. Cox, J. G. Louderback and L. A. Bloomfield. Phys Rev Lett., 71, 923
(1993)

\bibitem{be95}
G. Bertsch, N. Onishi and K. Yabana, Z Phys D, 34, 213 (1995)

\bibitem{be96}
G. Bertsch, N. Onishi and K. Yabana, Surf Rev Lett, 3, 435 (1996)

\bibitem{be94}
G.F. Bertsch and K. Yabana, Phys Rev A, 49, 1930 (1994)

\bibitem{du01}
P. Dugourd, I. Campagnon, F. Lepine, R. Antoine, D. Rayane and M. Broyer,
Chem Phys Lett., 336, 511 (2001)

\bibitem{web}
Program available at\\ 
http://gene.phys.washington.edu/$\sim$bertsch/adiabatic.f

\bibitem{typo} We note a typographical error in the formula as presented
in ref. \cite{be95}, eq. (2.23) of that reference.

\end{thebibliography}
\end{document}